\documentclass[journal]{IEEEtran}
\usepackage{bm}
\usepackage{cite}
\usepackage{amsmath,amssymb,amsfonts}
\usepackage{algorithm}
\usepackage{algpseudocode}
\usepackage{graphicx}
\usepackage{textcomp}
\usepackage{xcolor}
\usepackage{subfigure}
\allowdisplaybreaks[4]

\def\BibTeX{\rm B\kern-.05em{\sc i\kern-.025em b}\kern-.08emT\kern-.1667em\lower.7ex\hbox{E}\kern-.125emX}

\newtheorem{corollary}{Corollary}

\newtheorem{theorem}{Theorem}

\begin{document}

\title{Beamforming Design for Active RIS-Aided Over-the-Air Computation}

\author{Deyou Zhang, Ming Xiao, Mikael Skoglund, and H. Vincent Poor \vspace{-2em}

\thanks{D. Zhang, M. Xiao, and M. Skoglund are with the Division of Information Science and Engineering, KTH Royal Institute of Technology, Stockholm 10044, Sweden (email: \{deyou, mingx, skoglund\}@kth.se). H. V. Poor is with the Department of Electrical and Computer Engineering, Princeton University, Princeton, NJ 08544, USA (poor@princeton.edu).}
}

\maketitle

\begin{abstract}
Over-the-air computation (AirComp) is emerging as a promising technology for wireless data aggregation. However, its performance is hampered by users with poor channel conditions. To mitigate such a performance bottleneck, this paper introduces an active reconfigurable intelligence surface (RIS) into the AirComp system. Specifically, we begin by exploring the ideal RIS model and propose a joint optimization of the transceiver design and RIS configuration to minimize the mean squared error (MSE) between the target and estimated function values. To manage the resultant tri-convex optimization problem, we employ the alternating optimization (AO) technique to decompose it into three convex subproblems, each solvable optimally. Subsequently, we investigate two specific cases and analyze their respective asymptotic performance to reveal the superiority of the active RIS in mitigating the MSE relative to its passive counterpart. Lastly, we adapt our transceiver and RIS configuration design to account for the self-interference of the active RIS. To handle the resultant highly non-convex problem, we further devise a two-layer AO framework. Simulation results demonstrate the superiority of the active RIS in enhancing AirComp performance compared to its passive counterpart.
\end{abstract}

\begin{IEEEkeywords}
Over-the-air computation, active reconfigurable intelligent surface, wireless data aggregation.
\end{IEEEkeywords}

\IEEEpeerreviewmaketitle

\section{Introduction}
Wireless data aggregation (WDA) is envisioned as a common operation in 5G-and-beyond communications to support the Internet of Things (IoT), and it lays the foundation for numerous applications, e.g., distributed sensing, learning, and control \cite{GuangxuZhu-Magazine, ZhibinWang-Survey}. However, current WDA systems are often designed in a ``transmit-then-compute'' manner, separating communication and computation, which might encounter difficulty in accommodating a large number of IoT users under limited radio resources and stringent latency constraints. To address this concern, a novel technique termed over-the-air computation (AirComp) has been developed for WDA. By enabling concurrent data transmission from all IoT users over the same radio resources and exploiting the waveform superposition property of multiple access channels, AirComp turns the air into a computer to compute a specific function of these IoT users' data. Therefore, by integrating communication and computation, AirComp has the potential to achieve ultra-fast WDA even over massive IoT networks.

Recent research has extensively explored AirComp, particularly from the perspective of transceiver design \cite{ZhibinWang-Survey}. Specifically, \cite{XiaowenCao-OTA, WanchunLiu-OTA, XinZang-OTA} focused on the single-input single-output (SISO) configuration, investigating the optimal transceiver design for AirComp, which was later extended to multi-cluster scenarios in \cite{XiaowenCao-Multicell}. Moreover, \cite{LiChen-SIMO-UF} studied AirComp within the single-input multiple-output (SIMO) configuration, introducing a uniform-forcing transceiver design to manage non-uniform channel fading among users and generalizing such a methodology to multi-cluster scenarios in \cite{LiChen-SIMO-UF-Multicell}. To enable multi-function computation or multi-modal sensing, \cite{LiChen-MIMO-UF, Zhu-MIMO-DAB, Zhu-MIMO} investigated multiple-input multiple-output (MIMO) AirComp, with \cite{Zhu-MIMO} presenting a closed-form suboptimal solution for transmit and receive beamformers design. In addition to the traditional fully digital beamforming architecture, \cite{Zhai-Hybrid} and \cite{Xiao-Hybrid} proposed a hybrid analog-digital beamforming architecture for massive MIMO AirComp. As elucidated in these works, users with poor channel conditions dominate AirComp performance. Though massive MIMO technologies can significantly reduce computation errors, they entail substantial energy consumption and hardware complexity. Consequently, developing an economical yet efficient strategy to enhance AirComp performance remains crucially pertinent.

Reconfigurable intelligent surfaces (RISs) have emerged as an innovative technology to mitigate unfavorable channel conditions by re-configuring the radio propagation environment \cite{RuiZhang-RIS}. Compared to massive MIMO, RISs entail modest fabrication costs and energy consumption due to their use of passive reflective elements \cite{RuiZhang-RIS}. Owing to these advantages, RISs have permeated diverse applications, including cell-free communications \cite{ShaochengHuang-RIS, XinyingMa-RIS} and wireless federated learning \cite{My-TWC-FL}. RISs have also been introduced into AirComp systems \cite{Shitz-OTA-RIS, ZhibinWang-IoTJ, YMShi-OTA-RIS, Zhai-TwoStage, Zhai-STAR-RIS, Zhai-DRIS}. Specifically, \cite{Shitz-OTA-RIS} studied the advantages of deploying RISs for AirComp in a large-scale cloud radio access network. In \cite{ZhibinWang-IoTJ}, the authors integrated AirComp with energy harvesting and proposed leveraging RIS to enhance the efficacy of both operations. As the semi-definite relaxation algorithm was applied to optimize the beamforming matrix of the RIS in \cite{Shitz-OTA-RIS, ZhibinWang-IoTJ}, resulting in high computational complexity, \cite{YMShi-OTA-RIS} thereby developed a computationally efficient algorithm for RIS configuration design. To avoid excessive overhead caused by frequent channel state information estimation, \cite{Zhai-TwoStage} proposed a two-stage stochastic optimization strategy for RIS-aided AirComp systems. Moreover, to accommodate users distributed across the entire $360^{\circ}$ angular range, \cite{Zhai-STAR-RIS} advocated for the employment of a simultaneously transmitting and reflecting RIS (STAR-RIS). In this work, a joint optimization of the transceiver design, along with the reflective and refractive beamforming matrices of the STAR-RIS, was investigated to enhance AirComp performance.

Despite the growing popularity of RISs in AirComp-related research, the ``multiplicative fading'' effect substantially curtails its benefits. To mitigate such an inherent limitation, \cite{Zhai-DRIS} proposed deploying double RISs to assist AirComp, with one positioned near users and the other near the access point (AP). Alternatively, we can employ a novel RIS architecture, namely active RISs \cite{LinglongDai-ActiveRIS, YCLiang-ActiveRIS}, to enhance AirComp performance. Unlike its passive counterpart, the active RIS is capable of amplifying its reflected signals through reflection-type amplifiers integrated into its reflective elements, thereby overcoming the multiplicative fading effect. Given this advantage, active RISs have received remarkable attention \cite{LinglongDai-ActiveRIS, YCLiang-ActiveRIS, CunhuaPan-ActiveRIS-MEC, RuiZhang-ActiveRIS, CunhuaPan-ActiveRIS-Power}. Specifically, \cite{LinglongDai-ActiveRIS} explored the use of an active RIS to achieve substantial capacity gain for multi-user multiple-input single-output (MU-MISO) systems, while \cite{CunhuaPan-ActiveRIS-MEC} proposed deploying an active RIS to help minimize the computational latency for mobile edge computing systems. However, the corresponding transceiver and RIS configuration design approaches in these works cannot be directly applied to AirComp systems due to fundamentally different objectives, and to the best of our knowledge, employing active RISs to support AirComp remains unexplored in the open literature.

In what follows, we seek to fill this research gap by exploring an active RIS-aided AirComp system, wherein a multi-antenna AP aggregates data from multiple users using AirComp with the aid of an active RIS. Aligning with existing literature, we evaluate the performance of AirComp in terms of the mean squared error (MSE) between the target and estimated function values, and our objective is to develop a low-complexity approach for transceiver design and RIS configuration that minimizes the MSE. The main contributions of this paper are summarized as follows:

1) Firstly, we consider an ideal signal model for active RISs and formulate an MSE minimization problem, where each user's transmit equalization coefficient, the AP combining vector, and the beamforming matrix of the active RIS are jointly optimized. Given the non-convex objective function and coupled optimization variables, the resultant optimization problem is non-convex. To manage this difficulty, we establish an alternating optimization (AO) framework with a convergence guarantee to decompose the original problem into three subproblems, corresponding to the transmit equalization coefficients of users, the AP combining vector, and the beamforming matrix of the active RIS, respectively, and each of them can be efficiently solved.

2) Subsequently, we investigate two specific active RIS-aided AirComp systems: single-user (SU) SISO, in which the number of RIS elements is sufficiently large, and MU-SIMO, in which the number of AP antennas is equal to the number of RIS elements, both being sufficiently large. We analyze and compare their respective performances with those of passive RIS-aided systems, aiming to reveal the advantages of integrating active RISs into AirComp systems.

3) Furthermore, since the active RIS works in full-duplex mode, it incurs self-interference in practical systems. Consequently, we also extend the studied transceiver and RIS configuration design to scenarios involving self-interference of the active RIS. To handle the resultant highly non-convex problem, we employ the alternating direction method of multipliers (ADMM) and devise a two-layer AO framework to alternately optimize the transmit equalization coefficients of users, the AP combining vector, and the beamforming matrix of the active RIS.

4) Lastly, we carry out extensive simulations to demonstrate the superior performance of the active RIS-aided AirComp. Results show that the active RIS can yield a substantial performance improvement compared to its passive counterpart as long as its self-interference is not severe.

The remainder of this paper is organized as follows. In Section \ref{Sec-SM}, we introduce the ideal system model and formulate an optimization problem for transceiver and RIS configuration design. In Section \ref{Sec-TRR-Design}, we employ the AO technique to solve the formulated optimization problem. In Section \ref{Sec-PA}, we analyze the asymptotic performance of the considered active RIS-aided AirComp system and compare it to that of the passive RIS-aided one. In Section \ref{Sec-SI}, we extend the studied transceiver and RIS configuration design to account for the self-interference of the active RIS. Numerical results are provided in Section \ref{Sec-NR}, and we conclude this paper in Section \ref{Sec-CN}.

Throughout this paper, we use regular, bold lowercase, and bold uppercase letters to denote scalars, vectors, and matrices, respectively; $\mathcal R$ and $\mathcal C$ to denote real and complex number sets, respectively; $(\cdot)^*$, $(\cdot)^T$, and $(\cdot)^H$ to denote conjugate, transpose, and conjugate transpose, respectively. We use $\|\bm x\|$ to denote the $\ell_2$-norm of $\bm x$; $\|\bm X\|_{\rm F}$ to denote the Frobenius norm of $\bm X$; ${\rm diag}(\bm X)$ to denote the main diagonal of $\bm X$; ${\rm diag}(\bm x)$ or ${\rm diag}(\bm x^T)$ to denote a diagonal matrix with its diagonal entries specified by $\bm x$ or $\bm x^T$. We use $\bm I$ to denote the identity matrix; $\mathcal {CN}(\bm \mu, \bm \Sigma)$ to denote the complex Gaussian distribution with mean $\bm \mu$ and covariance matrix $\bm \Sigma$; $\mathbb{E}$ to denote the expectation operator; and $\odot$ to denote the Hadamard product.

\begin{figure}
\centering
\includegraphics[width = 6.4cm]{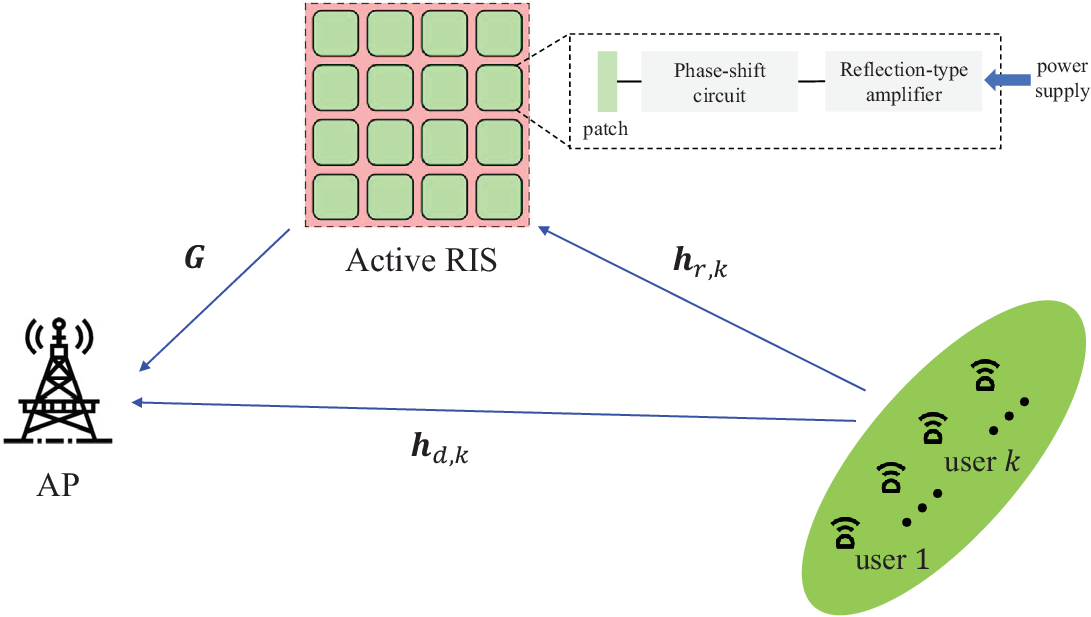}
\caption{Over-the-air computation via an active RIS.}\label{Fig-SM}
\end{figure}

\section{System Model and Problem Formulation}\label{Sec-SM}
\subsection{System Model}
We consider an MU-SIMO communication system consisting of $K$ single antenna users and an AP equipped with $M$ antennas. Let $s_k \in {\mathcal C}$ denote the data generated by user $k$, $\forall k \in \mathcal K \triangleq \{1, \cdots, K\}$. For the sake of simplicity, we assume $s_k$ has zero mean and unit variance, i.e., ${\mathbb E}[s_k] = 0$, and ${\mathbb E}[|s_k|^2] = 1$, $\forall k \in \mathcal K$, and $s_1, \cdots, s_K$ are uncorrelated with each other, i.e., ${\mathbb E}[s_k s^*_j] = 0$, $\forall k \ne j$. In this paper, we aim to recover the arithmetic mean of all users' data:
\begin{equation}
    s = \frac{1}{K}\sum\limits_{k=1}^K s_k,
\end{equation}
at AP by exploiting the superposition property of the wireless multiple-access channel.

To enhance the channel quality between users and AP, we propose deploying an $N$-element active RIS in the system, as shown in Fig. \ref{Fig-SM}. As a result, the equivalent channel between users and AP encompasses three links, i.e., the user-AP link, the user-RIS link, and the RIS-AP link, denoted by $\bm h_{d, k} \in {\cal C}^{M \times 1}$, $\bm h_{r, k} \in {\cal C}^{N \times 1}$, and $\bm G \in {\cal C}^{M \times N}$, respectively.

Referring to \cite{LinglongDai-ActiveRIS}, the reflected signal from the active RIS can be expressed as follows:
\begin{equation}
    \bm x_r = \bm \Phi \left(\sum\limits_{k=1}^K \bm h_{r, k} b_k s_k + \bm z_r\right),
\end{equation}
where $b_k$ is the transmit equalization coefficient of user $k$, $\forall k \in \mathcal K$, $\bm \Phi = {\rm diag}(\alpha_1 e^{j \omega_1}, \cdots, \alpha_N e^{j \omega_N})$ is the beamforming matrix of the active RIS, and $\bm z_r \sim {\cal CN}(\bm 0, \sigma_r^2 \bm I)$ is the thermal noise introduced at the active RIS. Due to the use of active components, $\alpha_n$, $\forall n \in {\cal N} \triangleq \{1, \cdots, N\}$, can be larger than one. Given $\bm x_r$, we can then express the received signal at AP as follows:
\begin{align}
    \bm y & = \sum\limits_{k=1}^K \bm h_{d, k} b_k s_k + \bm G \bm x_r + \bm z_a \nonumber \\[1ex]
    & = \sum\limits_{k=1}^K \left(\bm h_{d, k} + \bm G \bm \Phi \bm h_{r, k}\right) b_k s_k + \bm G \bm \Phi \bm z_r + \bm z_a,
\end{align}
where $\bm z_a \sim {\cal CN}(\bm 0, \sigma_a^2 \bm I)$ is the thermal noise at AP. Note that each user, as well as the active RIS, has its own maximum power constraint, given by 
\begin{align}
    {\mathbb E}[|b_k s_k|^2] = |b_k|^2 \le P_k, ~\forall k \in \mathcal K, ~~~~~~~~~~~~ \label{Eq-NodePowerCons} \\[2ex]
    {\mathbb E}[\|\bm x_r\|^2] = \sum\limits_{k=1}^K |b_k|^2 \|\bm \Phi \bm h_{r, k}\|^2 + \sigma_r^2 {\rm Tr}(\bm \Phi \bm \Phi^H) \le P_r, \label{Eq-RISPowerCons}
\end{align}
where $P_k$ and $P_r$ denote the maximum transmission power of user $k$ and the active RIS, respectively.

Let $\bm m \in {\cal C}^{M \times 1}$ denote the combining vector at AP; the estimated target variable is then given by
\begin{equation}
    \hat s = \bm m^H \bm y = \sum\limits_{k=1}^K \bm m^H \bm h_{e, k} b_k s_k + \bm m^H \left(\bm G \bm \Phi \bm z_r + \bm z_a\right),
\end{equation}
where $\bm h_{e, k} = \bm h_{d, k} + \bm G \bm \Phi \bm h_{r, k}$, $\forall k \in \mathcal K$, is termed the equivalent channel from user $k$ to AP.

\subsection{Problem Formulation}
Our objective is to minimize the distortion between the target variable and the estimated one, measured by the MSE as defined below:
\begin{align}\label{Eq-MSE}
    & \mathbb{MSE} \triangleq {\mathbb E}[|\hat s - s|^2] \\[2ex]
    & = {\mathbb E}\left[\left|\sum\limits_{k=1}^K \left(\bm m^H \bm h_{e, k} b_k - \frac{1}{K}\right) s_k + \bm m^H (\bm G \bm \Phi \bm z_r + \bm z_a)\right|^2\right] \nonumber \\[2ex]
    & = \sum\limits_{k=1}^K \left|\bm m^H \bm h_{e, k} b_k - \frac{1}{K}\right|^2 + \sigma_r^2 \|\bm m^H \bm G \bm \Phi\|^2 + \sigma_a^2 \|\bm m\|^2. \nonumber
\end{align}
To this end, we construct the following optimization problem:
\begin{subequations}\label{OP1}
    \begin{align}
        \min\limits_{\bm m, \bm b, \bm \Phi} &~ \mathbb{MSE}(\bm m, \bm b, \bm \Phi) \label{OP1-Obj} \\[2ex]
        \text{s.t.}~~&~ |b_k|^2 \le P_k, ~\forall k \in \mathcal K, \label{OP1-NodePowerCons} \\[2ex]
        &~ \sum\limits_{k=1}^K |b_k|^2 \|\bm \Phi \bm h_{r, k}\|^2 + \sigma_r^2 {\rm Tr}(\bm \Phi \bm \Phi^H) \le P_r, \label{OP1-RISPowerCons}
    \end{align}
\end{subequations}
where $\bm b \triangleq [b_1, \cdots, b_K]^T$.

\section{Alternating Optimization for Transceiver and RIS Configuration Design}\label{Sec-TRR-Design}
To address the coupling among $\bm m$, $\bm b$, and $\bm \Phi$ in \eqref{OP1}, we employ the AO technique, optimizing one variable at a time while holding the others fixed, as detailed below.

1) Optimization of $\bm m$: The associated optimization problem with respect to $\bm m$ is given by
\begin{equation}\label{Opt-m}
    \min\limits_{\bm m}~\mathbb{MSE},
\end{equation}
which is a least squares problem. The optimal $\bm m$ to \eqref{Opt-m} can be found by setting $\partial \mathbb{MSE} / \partial \bm m^*$ to zero, i.e.,
\begin{equation}
    \frac{\partial \mathbb{MSE}}{\partial \bm m^*} = \bm R \bm m - \frac{1}{K} \sum\limits_{k=1}^K \bm h_{e,k} b_k = \bm 0,
\end{equation}
which yields
\begin{equation}\label{Optimal-m}
    \bm m^{\star} = \bm R^{-1} \left(\frac{1}{K} \sum\limits_{k=1}^K \bm h_{e,k} b_k\right),
\end{equation}
where $\bm R = \sum\nolimits_{k=1}^K |b_k|^2 \bm h_{e, k} \bm h_{e,k}^H + \sigma_r^2 \bm G \bm \Phi \bm \Phi^{H} \bm G^H + \sigma_a^2 \bm I$. In addition, by substituting \eqref{Optimal-m} into \eqref{Eq-MSE}, we have
\begin{align}\label{Eq-MSE2}
    \mathbb{MSE} & = \frac{1}{K} - \frac{1}{K^2} \left(\sum\limits_{k=1}^K \bm h^H_{e, k} b^*_k\right) \bm R^{-1} \left(\sum\limits_{k=1}^K \bm h_{e, k} b_k\right) \\[1ex]
    & \overset{(a)}{<} \frac{1}{K}, \nonumber
\end{align}
where $(a)$ follows since $\bm R$ is a positive definite matrix.

2) Optimization of $\bm b$: The associated optimization problem with respect to $\bm b$ is formulated as follows:
\begin{subequations}\label{Opt-b}
    \begin{align}
        \min\limits_{\bm b} ~& \sum\limits_{k=1}^K \left|\bm m^H \bm h_{e, k} b_k - \frac{1}{K} \right|^2 \\[2ex]
        \text{s.t.}~~&~ |b_k|^2 \le P_k, ~\forall k \in {\cal K}, \\[2ex]
        &~ \sum\limits_{k=1}^K |b_k|^2 \|\bm \Phi \bm h_{r, k}\|^2 + \sigma_r^2 {\rm Tr}(\bm \Phi \bm \Phi^H) \le P_r.
    \end{align}
\end{subequations}
It is observed that \eqref{Opt-b} is a quadratically constrained quadratic program (QCQP), and off-the-shelf solvers such as CVX can be used to efficiently solve this problem.

3) Optimization of $\bm \Phi$: The associated optimization problem with respect to $\bm \Phi$ is formulated as follows:
\begin{subequations}\label{Opt-Phi}
\begin{align}
    \min\limits_{\bm \Phi} &~ \sum\limits_{k=1}^K \left|\bm m^{H} \bm h_{e, k} b_k - \frac{1}{K}\right|^2 + \sigma_r^2 \|\bm m^H \bm G \bm \Phi\|^2 \label{Opt-Phi-Obj} \\[2ex]
    \text{s.t.} &~ \sum\limits_{k=1}^K |b_k|^2 \|\bm \Phi \bm h_{r, k}\|^2 + \sigma_r^2 {\rm Tr}(\bm \Phi \bm \Phi^H) \le P_r. \label{Opt-Phi-PowerCons}
\end{align}
\end{subequations}
By introducing $\bm \phi \triangleq {\rm diag}(\bm \Phi)$ and applying the property of $\bm \Phi \bm x = {\rm diag}(\bm x) \bm \phi$, we can equivalently transform \eqref{Opt-Phi} into (see Appendix \ref{App-Opt-Phi}) 
\begin{subequations}\label{Opt-Phi2}
\begin{align}
    \min\limits_{\bm \phi} &~ \bm \phi^H \bm A_0 \bm \phi - \bm \phi^H \bm v_0 - \bm v_0^H \bm \phi \\[2ex]
    \text{s.t.} &~ \bm \phi^H \bm B_0 \bm \phi \le P_r,
\end{align}
\end{subequations}
where $\bm A_0$, $\bm B_0$, and $\bm v_0$ are respectively given by
\begin{align*}
    & \bm A_0 = \sum\limits_{k=1}^K \bm a_k \bm a_k^H  + \sigma_r^2 {\rm diag}(\bm m^H \bm G) {\rm diag}(\bm G^H \bm m), \\[2ex]
    & \bm B_0 = \sum\limits_{k = 1}^K |b_k|^2 {\rm diag}(\bm h^*_{r,k}) {\rm diag}(\bm h_{r,k}) + \sigma_r^2 \bm I, \\[2ex]
    & \bm v_0 = \sum\limits_{k=1}^K \left(\frac{1}{K} - \bm m^{H} \bm h_{d, k} b_k\right) \bm a_k, \\[2ex]
    & \bm a_k = {\rm diag}(\bm h^{*}_{r,k} b_k^{*}) \bm G^H \bm m, ~\forall k \in \mathcal K.
\end{align*}
It is observed that \eqref{Opt-Phi2} is a standard QCQP and can be solved optimally using the Karush-Kuhn-Tucker (KKT) conditions, as detailed below.

First of all, the Lagrangian associated with \eqref{Opt-Phi2} is defined as follows:
\begin{equation}\label{Opt-Phi2-Lagrangian}
    L_0 = \bm \phi^H \bm A_0 \bm \phi - \bm \phi^H \bm v_0 - \bm v_0^H \bm \phi + \lambda_0 (\bm \phi^H \bm B_0 \bm \phi - P_r),
\end{equation}
where $\lambda_0 \ge 0$ is the Lagrange multiplier. The KKT conditions of \eqref{Opt-Phi2-Lagrangian} are given by
\begin{subequations}
    \begin{align}
        \frac{\partial L_0}{\partial \bm \phi^*} = \left(\bm A_0 + \lambda_0 \bm B_0 \right) \bm \phi - \bm v_0 = \bm 0, \label{Opt-Phi2-KKT-phi} \\[2ex]
        \lambda_0 (\bm \phi^H \bm B_0 \bm \phi - P_r) = 0, \label{Opt-Phi2-KKT-Slackness} \\[2ex]
        \bm \phi^H \bm B_0 \bm \phi = P_r. \label{Opt-Phi2-KKT-PowerCons}
     \end{align}
\end{subequations}
From \eqref{Opt-Phi2-KKT-phi}, we can obtain the optimal $\bm \phi$ as follows:
\begin{equation}\label{Optimal-phi}
    \bm \phi^{\star} = \left(\bm A_0 + \lambda_0 \bm B_0 \right)^{-1} \bm v_0,
\end{equation}
where the nonnegative Lagrange multiplier $\lambda_0$ should be chosen to satisfy \eqref{Opt-Phi2-KKT-Slackness} and \eqref{Opt-Phi2-KKT-PowerCons}. With \eqref{Optimal-phi}, we can verify that $(\bm \phi^{\star})^H \bm B_0 \bm \phi^{\star}$ is a decreasing function of $\lambda_0$. Moreover, it can be shown that
\begin{equation}\label{lambda-bounds}
    0 \le \lambda_0 < \sqrt{\frac{\bm v_0^H \bm B_0^{-1} \bm v_0}{P_r}}.
\end{equation}
Therefore, we can search for $\lambda_0$ using the bisection search method within the bounds on $\lambda_0$ in \eqref{lambda-bounds}.

The proposed transceiver and RIS configuration design approach for the MU-SIMO AirComp system is outlined in \textbf{Algorithm \ref{Alg-ID}}. Additionally, the convergence of \textbf{Algorithm \ref{Alg-ID}} is elucidated in the subsequent theorem. 
\begin{theorem}
    \textbf{Algorithm \ref{Alg-ID}} converges to a locally optimal point of problem \eqref{OP1} after several iterations.
\end{theorem}
\begin{IEEEproof}
To prove the convergence of \textbf{Algorithm \ref{Alg-ID}}, we introduce superscript $t$ as the iteration index, e.g., $\bm m^t$ denotes the combining vector at the end of the $t$-th iteration round. Then, \textbf{Algorithm \ref{Alg-ID}} converges as
\begin{align*}
    \mathbb{MSE}(\bm m^t, \bm b^t, \bm \Phi^t) & \overset{(a)}{\ge} \mathbb{MSE}( \bm m^{t+1}, \bm b^t, \bm \Phi^t) \\[2ex]
    & \overset{(b)}{\ge} \mathbb{MSE}(\bm m^{t+1}, \bm b^{t+1}, \bm \Phi^t) \\[2ex]
    & \overset{(c)}{\ge} \mathbb{MSE}(\bm m^{t+1}, \bm b^{t+1}, \bm \Phi^{t+1}),
\end{align*}
where $(a)$, $(b)$, and $(c)$ follow since the updates of $\bm m$, $\bm b$, and $\bm \Phi$ are the optimal solutions to \eqref{Opt-m}, \eqref{Opt-b}, and \eqref{Opt-Phi}, respectively. Since $\mathbb{MSE}(\bm m^t, \bm b^t, \bm \Phi^t)$ is monotonically non-increasing in each iteration and its value is lower bounded by zero, we prove that \textbf{Algorithm \ref{Alg-ID}} will converge to a local optimum of problem \eqref{OP1} after several iterations.
\end{IEEEproof}

\begin{algorithm}[!htbp]
    \caption{Pseudo-Code for the Proposed Transceiver and RIS Configuration Design}
    \begin{algorithmic}[1]
        \State Initialize $\bm b$ and $\bm \Phi$ to ensure that they satisfy the power constraints in \eqref{OP1-NodePowerCons} and \eqref{OP1-RISPowerCons};
        \While{not converge}
        \State Update $\bm m$ through \eqref{Optimal-m};
        \State Update $\bm b$ through solving \eqref{Opt-b};
        \State Compute $\bm \phi$ through \eqref{Optimal-phi};
        \State Update $\bm \Phi$ by $\bm \Phi = {\rm diag}(\bm \phi)$.
        \EndWhile
    \end{algorithmic}\label{Alg-ID}
\end{algorithm}

\section{Performance Analysis}\label{Sec-PA}
In this section, our goal is to evaluate the asymptotic performance of the active RIS-aided AirComp system and compare it to its passive RIS-aided counterpart. To make the analysis tractable and yield insightful results, two special cases are examined\footnote{Analyzing the asymptotic performance of the active RIS-aided AirComp system under general settings can be extremely challenging, if not possible.}: SU-SISO with $N \to \infty$, and MU-SIMO with $M = N \to \infty$.

\subsection{SU-SISO System}\label{Sec-SISO-PA}
In alignment with \cite{RuiZhang-RIS} and \cite{LinglongDai-ActiveRIS}, we temporarily adopt the following settings: 1) A SU-SISO system is considered, i.e., $K = M = 1$; 2) The direct link between user and AP is sufficiently weak to be ignored, i.e., $h_d = 0$; 3) Each RIS element maintains the same amplification factor, i.e., $\alpha_n = \alpha$, $\forall n \in \mathcal N$; and 4) Both the user-RIS channel $\bm h_r$ and the RIS-AP channel $\bm g^H$ follow a Rayleigh distribution, i.e., $\bm h_r \sim {\cal CN}(\bm 0, \rho^2_r \bm I)$, and $\bm g \sim {\cal CN}(\bm 0, \rho^2_g \bm I)$. With these simplifications, we derive a closed-form asymptotic MSE, as detailed in the subsequent theorem.

\begin{theorem}\label{Theorem-SISO-ActiveRIS}
    Letting $N \to \infty$, the asymptotic MSE for the active RIS-aided SU-SISO system is given by
    \begin{equation}\label{MSE-SISO-ActiveRIS}
        \mathbb{MSE}^{\rm su-siso}_{\rm active} \approx \frac{16}{\pi^2 N} \frac{\sigma_r^2 \sigma_a^2 + P_0 \rho^2_r \sigma_a^2 + P_r \rho^2_g \sigma_r^2}{P_r P_0 \rho_r^2 \rho_g^2},
    \end{equation}
    where $P_0$ is the maximum transmission power of the user.
\end{theorem}
\begin{IEEEproof}
    Refer to Appendix \ref{Analysis-SU-SISO}.
\end{IEEEproof}

\emph{Remark 1}: The asymptotic MSE in \eqref{MSE-SISO-ActiveRIS} is equal to the inverse of the asymptotic SNR in [\citenum{LinglongDai-ActiveRIS}, Lemma 2].

For comparison, under the same settings, we also provide the asymptotic MSE for the passive RIS-aided SU-SISO system in the subsequent theorem.
\begin{theorem}\label{Theorem-SISO-PassiveRIS}
     Letting $N \to \infty$, the asymptotic MSE for the passive RIS-aided SU-SISO system is given by
    \begin{equation}\label{MSE-SISO-PassiveRIS}
        \mathbb{MSE}^{\rm su-siso}_{\rm passive} \approx \frac{16}{\pi^2 N^2} \frac{\sigma_a^2}{\tilde P_0 \rho_r^2 \rho_g^2},
    \end{equation}
    where $\tilde P_0$ is the maximum transmission power of the user in the passive RIS-aided system.
\end{theorem}
\begin{IEEEproof}
     Refer to Appendix \ref{Analysis-SU-SISO2}.
\end{IEEEproof}

It can be observed from \eqref{MSE-SISO-ActiveRIS} that the asymptotic MSE for the active RIS-aided SU-SISO system is inversely proportional to $N$, whereas the asymptotic MSE for its passive RIS-aided counterpart is inversely proportional to $N^2$. Consequently, we can deduce that $\mathbb{MSE}^{\rm su-siso}_{\rm passive}$ will be lower than $\mathbb{MSE}^{\rm su-siso}_{\rm active}$ when $N$ is sufficiently large. However, this may not hold in scenarios where $N$ is not that large. The rationale behind this is that the nominator of \eqref{MSE-SISO-ActiveRIS} can be substantially smaller than that of \eqref{MSE-SISO-PassiveRIS}, due to the multiplicative terms composed of path losses and noise power, namely, $P_0 \rho^2_r \sigma_a^2$, $\sigma_r^2 \sigma_a^2$, and $P_r \rho^2_g \sigma_r^2$. The subsequent corollary delineates the condition under which the passive RIS outperforms its active counterpart.

\begin{corollary}
    To ensure that $\mathbb{MSE}^{\rm su-siso}_{\rm passive} \le \mathbb{MSE}^{\rm su-siso}_{\rm active}$, the required number of RIS elements $N$ has to satisfy
    \begin{equation}
        N \ge N_{th} = \frac{P_0}{\tilde P_0} \frac{P_r \sigma_a^2}{\sigma_r^2 \sigma_a^2 + P_0 \rho^2_r \sigma_a^2 + P_r \rho^2_g \sigma_r^2}.
    \end{equation}
\end{corollary}
\begin{IEEEproof}
    By solving $\mathbb{MSE}^{\rm su-siso}_{\rm passive} \le \mathbb{MSE}^{\rm su-siso}_{\rm active}$, we can readily prove this corollary.
\end{IEEEproof}

\emph{Remark 2:} Given $\sigma_a^2 = \sigma_r^2 = -100$ dB, $\rho_r^2 = \rho_g^2 = -70$ dB, $P_0 = P_r = 1$ W, $\tilde P_0 = 2$ W, we calculate $N_{th}$ to be $2.5 \times 10^6$, a value that poses significant challenges for realization with current technology.

\subsection{MU-SIMO System}\label{Sec-SIMO-PA}
In order to make the analysis tractable, we temporarily adopt the following settings: 1) The number of AP antennas is equal to the number of RIS elements and sufficiently large, outnumbering the users, i.e., $K \le M = N \to \infty$; 2) The direct links between users and AP are weak enough to be ignored, i.e., $\bm h_{d,k} = \bm 0$, $\forall k \in \mathcal K$; 3) All the $N$ RIS elements maintain the same amplification factor, i.e., $\alpha_n = \alpha$, $\forall n \in {\cal N}$, with their phase shifts uniformly and randomly distributed between $(0, 2\pi)$; and 4) Both the user-RIS channel $\bm h_{r,k}$, $\forall k \in \mathcal K$, and the RIS-AP channel $\bm G$ follow a Rayleigh distribution, i.e., the entries in $\bm h_{r,k}$, $\forall k \in \mathcal K$, and $\bm G$ are independent and identically distributed, following ${\cal CN}(0, \rho^2_r)$ and ${\cal CN}(0, \rho^2_g)$, respectively. With these simplifications, we have the following useful properties.
\begin{subequations}
\begin{align}
    \frac{1}{N^2} \Bigg(\bm h^H_{e,k} \bm h_{e,k}\Bigg) & = \frac{1}{N^2} \Bigg(\bm h^H_{r,k} \bm \Phi^H \bm G^H \bm G \bm \Phi \bm h_{r,k}\Bigg) \nonumber \\[2ex]
    & \overset{(a)}{\approx} \frac{\rho^2_g}{N} \Bigg(\bm h^H_{r,k} \bm \Phi^H \bm \Phi \bm h_{r,k}\Bigg) \nonumber \\[2ex]
    & = \frac{\alpha^2 \rho^2_g}{N} \Bigg(\bm h^H_{r,k} \bm h_{r,k}\Bigg) \nonumber \\[2ex]
    & \overset{(b)}{\approx} \alpha^2 \rho^2_g \rho^2_r, ~\forall k \in \mathcal K, \label{Channel-Property1} \\[2ex]
    \frac{1}{N^2} \Bigg(\bm h^H_{e,k} \bm h_{e,k'}\Bigg) & = \frac{1}{N^2} \Bigg(\bm h^H_{r,k} \bm \Phi^H \bm G^H \bm G \bm \Phi \bm h_{r,k'}\Bigg) \nonumber \\[2ex]
    & \approx \frac{\rho^2_g}{N}~\Bigg(\bm h^H_{r,k} \bm \Phi^H \bm \Phi \bm h_{r,k'}\Bigg) \nonumber \\[2ex]
    & = \frac{\alpha^2 \rho^2_g}{N} \Bigg(\bm h^H_{r,k} \bm h_{r,k'}\Bigg) \nonumber \\[2ex]
    & \overset{(c)}{\approx} 0, ~\forall k, k' \in \mathcal K, \label{Channel-Property2}
\end{align}
\end{subequations}
where $(a)$, $(b)$, and $(c)$ follow due to the following properties:
\begin{subequations}
\begin{align}
    & \lim\limits_{N \to \infty} \frac{1}{N} \Bigg(\bm h^H_{r,k} \bm h_{r,k}\Bigg) = \rho^2_r, ~\forall k \in \mathcal K, \\[1ex]
    & \lim\limits_{N \to \infty} \frac{1}{N} \Bigg(\bm h^H_{r,k} \bm h_{r,k'}\Bigg) = 0, ~\forall k, k' \in \mathcal K, \\[1ex]
    & \lim\limits_{N \to \infty} \frac{1}{N} \Bigg(\bm G^H \bm G\Bigg) = \lim\limits_{N \to \infty} \frac{1}{N} \Bigg(\bm G \bm G^H\Bigg) = \rho^2_g \bm I. \label{Channel-Property3}
\end{align}
\end{subequations}
Subsequently, by leveraging \eqref{Channel-Property1}, \eqref{Channel-Property2}, and \eqref{Channel-Property3}, we derive a closed-form asymptotic MSE for the active RIS-aided MU-SIMO system, detailed in the theorem below.

\begin{theorem}\label{Theorem-SIMO-ActiveRIS}
    By letting $K \le M = N \to \infty$, the asymptotic MSE for the active RIS-aided MU-SIMO system is given by
    \begin{equation}\label{MSE-SIMO-ActiveRIS}
        \mathbb{MSE}^{\rm mu-simo}_{\rm active} \approx \frac{1}{K N} \frac{\sigma_r^2 \sigma_a^2 + K P_0 \rho^2_r \sigma_a^2 + P_r \rho^2_g \sigma_r^2}{P_r P_0 \rho_r^2 \rho_g^2},
    \end{equation}
    where $P_0$ is the maximum transmission power of each user, i.e., $P_k = P_0$, $\forall k \in \mathcal K$.
\end{theorem}
\begin{IEEEproof}
    Refer to Appendix \ref{Analysis-MU-SIMO}.
\end{IEEEproof}

Similar to the SU-SISO case, we also provide the asymptotic MSE for the passive RIS-aided MU-SIMO system in the following theorem for comparison.
\begin{theorem}\label{Theorem-SIMO-PassiveRIS}
    By letting $M = N \to \infty$, the asymptotic MSE for the passive RIS-aided MU-SIMO system is given by
    \begin{equation}\label{MSE-SIMO-PassiveRIS}
        \mathbb{MSE}^{\rm mu-simo}_{\rm passive} \approx \frac{1}{K N^2} \frac{\sigma_a^2}{\tilde P_0 \rho_r^2 \rho_g^2},
    \end{equation}
    where $\tilde P_0$ is the maximum transmission power of each user in the passive RIS-aided system.
\end{theorem}
\begin{IEEEproof}
     Refer to Appendix \ref{Analysis-MU-SIMO2}.
\end{IEEEproof}

Furthermore, the ensuing corollary delineates the condition under which the passive RIS-aided system outperforms its active counterpart.

\begin{corollary}
    To ensure that $\mathbb{MSE}^{\rm mu-simo}_{\rm passive} \le \mathbb{MSE}^{\rm mu-simo}_{\rm active}$, the required number of RIS elements $N$ has to satisfy
    \begin{equation}
        N \ge \frac{P_0}{\tilde P_0} \frac{P_r \sigma_a^2}{\sigma_r^2 \sigma_a^2 + K P_0 \rho^2_r \sigma_a^2 + P_r \rho^2_g \sigma_r^2}.
    \end{equation}
\end{corollary}
\begin{IEEEproof}
    By solving $\mathbb{MSE}^{\rm mu-simo}_{\rm passive} \le \mathbb{MSE}^{\rm mu-simo}_{\rm active}$, we can readily prove this corollary.
\end{IEEEproof}

\section{Self-Interference Suppression}\label{Sec-SI}
In this section, we extend the studied transceiver and RIS configuration design to scenarios involving self-interference of the active RIS.

\subsection{Self-Interference Modeling}
According to \cite{LinglongDai-ActiveRIS}, the reflected signal from the active RIS in the presence of self-interference can be expressed as follows:
\begin{align}\label{Eq-SI-Theta}
    \bm x_r & = (\bm I - \bm \Phi \bm H)^{-1} \bm \Phi \left(\sum\limits_{k=1}^K \bm h_{r, k} b_k s_k + \bm z_r\right) \nonumber \\[2ex]
    & \overset{(a)}{\approx} (\bm I + \bm \Phi \bm H) \bm \Phi \left(\sum\limits_{k=1}^K \bm h_{r, k} b_k s_k + \bm z_r\right),
\end{align}
where $\bm H \in {\cal C}^{N \times N}$ denotes the self-interference channel, and $(a)$ is due to $(\bm I - \bm \Phi \bm H)^{-1} \approx \bm I + \bm \Phi \bm H$ when all elements in $\bm H$ are small. In particular, when all elements in $\bm H$ are zero, it is observed that both $(\bm I - \bm \Phi \bm H)^{-1} \bm \Phi$ and $(\bm I + \bm \Phi \bm H) \bm \Phi$ will reduce to $\bm \Phi$.

\subsection{Problem Formulation}
To incorporate the self-interference of the active RIS into the transceiver and RIS configuration design, an intuitive way is to replace $\bm \Phi$ in \eqref{OP1} with $(\bm I + \bm \Phi \bm H) \bm \Phi$. Since this operation does not impact the optimizations of $\bm m$ and $\bm b$, we only focus on optimizing $\bm \Phi$. The corresponding optimization problem is expressed as follows:
\begin{subequations}\label{Opt-Theta}
\begin{align}
    \min\limits_{\bm \Phi} &~ f_1(\bm \Phi) \\[2ex]
    \text{s.t.} &~ g_1(\bm \Phi) \le P_r,
\end{align}
\end{subequations}
where $f_1(\bm \Phi)$ and $g_1(\bm \Phi)$ are respectively given by
\begin{align*}
    f_1(\bm \Phi) = & \sum\limits_{k=1}^K \left|\bm m^{H} (\bm h_{d, k} + \bm G (\bm I + \bm \Phi \bm H) \bm \Phi \bm h_{r, k}) b_k - \frac{1}{K}\right|^2 \\[2ex]
    & + \sigma_r^2 \|\bm m^{H} \bm G (\bm I + \bm \Phi \bm H) \bm \Phi\|^2, \\[2ex]
    g_1(\bm \Phi) = & \sum\limits_{k=1}^K |b_k|^2 \|(\bm I + \bm \Phi \bm H) \bm \Phi \bm h_{r, k}\|^2 \\[2ex]
    & + \sigma_r^2 {\rm Tr}((\bm I + \bm \Phi \bm H) \bm \Phi \bm \Phi^H (\bm I + \bm \Phi \bm H)^H).
\end{align*}

\subsection{RIS Configuration Design}
To handle \eqref{Opt-Theta}, we employ ADMM, introducing an auxiliary variable $\bm {\tilde \Phi} = \bm \Phi$ and reformulating problem \eqref{Opt-Theta} as follows:
\begin{subequations}\label{Opt-Theta2}
\begin{align}
    \min\limits_{\bm \Phi, \bm {\tilde \Phi}} &~ f_2(\bm \Phi, \bm {\tilde \Phi}) + \tau \|\bm {\tilde \Phi} - \bm \Phi\|^2_{\rm F} \label{Opt-Theta2-Obj} \\[2ex]
    \text{s.t.} &~ g_2(\bm \Phi, \bm {\tilde \Phi}) \le P_r, \label{Opt-Theta2-PowerCons}
\end{align}
\end{subequations}
where $f_2(\bm \Phi, \bm {\tilde \Phi})$ and $g_2(\bm \Phi, \bm {\tilde \Phi})$ are respectively given by
\begin{align*}
    f_2(\bm \Phi, \bm {\tilde \Phi}) = & \sum\limits_{k=1}^K \left|\bm m^{H} (\bm h_{d, k} + \bm G (\bm I + \bm {\tilde \Phi} \bm H) \bm \Phi \bm h_{r, k}) b_k - \frac{1}{K}\right|^2 \\[2ex]
    & + \sigma_r^2 \|\bm m^{H} \bm G (\bm I + \bm {\tilde \Phi} \bm H) \bm \Phi\|^2, \\[2ex]
    g_2(\bm \Phi, \bm {\tilde \Phi}) = & \sum\limits_{k=1}^K |b_k|^2 \|(\bm I + \bm {\tilde \Phi} \bm H) \bm \Phi \bm h_{r, k}\|^2 \\[2ex]
    & + \sigma_r^2 {\rm Tr}((\bm I + \bm {\tilde \Phi} \bm H) \bm \Phi \bm \Phi^H (\bm I + \bm {\tilde \Phi} \bm H)^H).
\end{align*}
Note that $\tau$ in \eqref{Opt-Theta2-Obj} is a tuning parameter, and it can be proven that \eqref{Opt-Theta2} is equivalent to \eqref{Opt-Theta} when $\tau \to \infty$. In the sequel, we recall the AO technique and optimize $\bm \phi \triangleq {\rm diag}(\bm \Phi)$ and $\bm {\tilde \phi} \triangleq {\rm diag}(\bm {\tilde \Phi})$ alternately until $\bm \phi = \bm {\tilde \phi}$ is achieved.

1) Optimization of $\bm \phi$: Via some mathematical manipulations to \eqref{Opt-Theta2}, we formulate an optimization problem with respect to $\bm \phi$ as follows (see Appendix \ref{App-Opt-Theta}):
\begin{subequations}\label{Opt-Theta3}
\begin{align}
    \min\limits_{\bm \phi} &~ \bm \phi^H \bm A_1 \bm \phi - \bm \phi^H \bm v_1 - \bm v_1^H \bm \phi \\[2ex]
    \text{s.t.} &~ \bm \phi^H \bm B_1 \bm \phi \le P_r,
\end{align}
\end{subequations}
where $\bm A_1$, $\bm B_1$, and $\bm v_1$ are respectively given by
\begin{align*}
    & \bm A_1 = \sum\limits_{k=1}^K \bm {\bar a}_k \bm {\bar a}_k^H + \tau \bm I \\[2ex]
    &~~~~~~~ + \sigma_r^2 {\rm diag}(\bm m^H \bm G \bm \Omega) {\rm diag}(\bm \Omega^H \bm G^H \bm m), \\[2ex]
    & \bm B_1 = \sum\limits_{k=1}^K |b_k|^2 {\rm diag}(\bm h^*_{r, k}) \bm \Omega^H \bm \Omega {\rm diag}(\bm h_{r, k}) \\[2ex]
    &~~~~~~~ + \sigma_r^2 ((\bm \Omega^H \bm \Omega) \odot \bm I), \\[2ex]
    & \bm \Omega = \bm I + \bm {\tilde \Phi} \bm H, \\[2ex]
    & \bm v_1 = \sum\limits_{k=1}^K \left(\frac{1}{K} - \bm m^{H} \bm h_{d, k} b_k\right) \bm {\bar a}_k + \tau \bm {\tilde \Phi}, \\[2ex]
    & \bm {\bar a}_k = {\rm diag}(\bm h^{*}_{r,k} b_k^{*}) \bm \Omega^H \bm G^H \bm m,~\forall k \in {\cal K}.
\end{align*}
Given that \eqref{Opt-Theta3} exhibits the same form as \eqref{Opt-Phi2}, the optimal $\bm \phi$ can be obtained using the same method as applied in \eqref{Opt-Phi2}. For brevity, we omit the details herein and only provide the optimal solution as follows:
\begin{equation}\label{Optimal-Theta}
    \bm \phi^{\star} = \left(\bm A_1 + \lambda_1 \bm B_1\right)^{-1} \bm v_1,
\end{equation}
where $\lambda_1$ is the associated Lagrange multiplier, and it can be determined through the bisection search method within the following bounds:
\begin{equation}
    0 \le \lambda_1 < \sqrt{\frac{\bm v_1^H \bm B_1^{-1} \bm v_1}{P_r}}.
\end{equation}

2) Optimization of $\bm {\tilde \phi}$: The associated optimization problem with respect to $\bm {\tilde \phi}$ is formulated as follows (see Appendix \ref{App-Opt-Theta2}):
\begin{subequations}\label{Opt-Theta4}
\begin{align}
    \min\limits_{\bm {\tilde \phi}} &~ \bm {\tilde \phi}^H \bm A_2 \bm {\tilde \phi} - \bm {\tilde \phi}^H \bm v_2 - \bm v_2^H \bm {\tilde \phi} \\[2ex]
    \text{s.t.} &~ \bm {\tilde \phi}^H \bm B_2 \bm {\tilde \phi} + \bm q^H \bm {\tilde \phi} + \bm {\tilde \phi}^H \bm q + {\rm Tr}(\bm D) \le P_r,
\end{align}
\end{subequations}
where $\bm A_2$, $\bm B_2$, $\bm D$, $\bm v_2$, and $\bm q$ are respectively given by
\begin{align*}
    & \bm A_2 = \sum\limits_{k=1}^K \bm {\tilde a}_k \bm {\tilde a}_k^H + \tau \bm I \nonumber \\[2ex]
    &~~~~~~~ + \sigma_r^2 {\rm diag}(\bm G^H \bm m) \bm H^* \bm \Phi^H \bm \Phi \bm H^T {\rm diag}(\bm m^H \bm G), \\[2ex]
    & \bm B_2 = (\bm H \bm D \bm H^H) \odot \bm I, \\[2ex]
    & \bm D  = \bm \Phi \left(\sum\limits_{k=1}^K |b_k|^2 \bm h_{r, k} \bm h^H_{r, k} + \sigma_r^2 \bm I\right) \bm \Phi^H, \\[2ex]
    & \bm v_2 = \sum\limits_{k=1}^K \left[\frac{1}{K} - \bm m^{H} (\bm h_{d, k} + \bm G \bm \Phi \bm h_{r,k}) b_k\right] \bm {\tilde a}_k + \tau {\bm \Phi} \\[2ex]
    &~~~~~~ - \sigma_r^2 {\rm diag}(\bm m^H \bm G \bm \Phi \bm \Phi^H \bm H^H) \bm G^H \bm m, \\[2ex]
    & \bm q = {\rm diag}(\bm D \bm H^H), \\[2ex]
    & \bm {\tilde a}_k = {\rm diag}(\bm H^{*} \bm \Phi^{*} \bm h^{*}_{r,k} b_k^{*}) \bm G^H \bm m, ~\forall k \in \mathcal K.
\end{align*}
It is observed that \eqref{Opt-Theta4} is a QCQP, and we can also employ the KKT conditions to solve it, as detailed below.

First of all, the Lagrangian associated with \eqref{Opt-Theta4} is expressed as follows:
\begin{align}\label{Opt-Theta4-Lagrangian}
    L_2 & = \bm {\tilde \phi}^H \bm A_2 \bm {\tilde \phi} - \bm {\tilde \phi}^H \bm v_2 - \bm v_2^H \bm {\tilde \phi} \nonumber \\[2ex]
    & + \lambda_2 (\bm {\tilde \phi}^H \bm B_2 \bm {\tilde \phi} + \bm q^H \bm {\tilde \phi} + \bm {\tilde \phi}^H \bm q + {\rm Tr}(\bm D) - P_r),
\end{align}
where $\lambda_2 \ge 0$ is the Lagrange multiplier. The KKT conditions of \eqref{Opt-Theta4-Lagrangian} are given by
\begin{subequations}
    \begin{align}
        \frac{\partial L_2}{\partial \bm {\tilde \phi}^*} = \left(\bm A_2 + \lambda_2 \bm B_2 \right) \bm {\tilde \phi} - \bm v_2 + \lambda_2 \bm q = \bm 0, \label{Opt-Theta4-KKT-phi} \\[1ex]
        \lambda_2 (\bm {\tilde \phi}^H \bm B_2 \bm {\tilde \phi} + \bm q^H \bm {\tilde \phi} + \bm {\tilde \phi}^H \bm q + {\rm Tr}(\bm D) - P_r) = 0, \label{Opt-Theta4-KKT-Slackness} \\[2ex]
        \bm {\tilde \phi}^H \bm B_2 \bm {\tilde \phi} + \bm q^H \bm {\tilde \phi} + \bm {\tilde \phi}^H \bm q + {\rm Tr}(\bm D) = P_r. \label{Opt-Theta4-KKT-PowerCons}
     \end{align}
\end{subequations}
From \eqref{Opt-Theta4-KKT-phi}, we obtain the optimal $\bm {\tilde \phi}$ as follows:
\begin{equation}\label{Optimal-TildeTheta}
    \bm {\tilde \phi}^{\star} = \left(\bm A_2 + \lambda_2 \bm B_2 \right)^{-1} (\bm v_2 - \lambda_2 \bm q),
\end{equation}
where the nonnegative Lagrange multiplier $\lambda_2$ can be determined through the one-dimensional grid search to satisfy \eqref{Opt-Theta4-KKT-Slackness} and \eqref{Opt-Theta4-KKT-PowerCons}.

\subsection{Overall Algorithm}
The proposed RIS beamforming design in the presence of RIS self-interference is outlined in \textbf{Algorithm \ref{Alg-SI-Suppress}}. Additionally, the whole procedure for transceiver and RIS configuration design considering RIS self-interference is summarized in \textbf{Algorithm \ref{Alg-SI}}. Furthermore, the convergence of \textbf{Algorithm \ref{Alg-SI-Suppress}} and \textbf{Algorithm \ref{Alg-SI}} can be proved analogously to \textbf{Algorithm \ref{Alg-ID}}.

\begin{algorithm}[!htbp]
    \caption{Pseudo-Code for the RIS Beamforming Design Considering RIS Self-Interference}
    \begin{algorithmic}[1]
        \State Initialize $\bm {\tilde \phi}$, and $\tau$;
        \While{not converge}
        \State Compute $\bm \phi$ through \eqref{Optimal-Theta};
        \State Compute $\bm {\tilde \phi}$ through \eqref{Optimal-TildeTheta};
        \State Update $\tau$ by $\tau \leftarrow 1.1 \times \tau$;
        \EndWhile
        \State Update $\bm \Phi$ by $\bm \Phi = {\rm diag}(\bm \phi)$.
    \end{algorithmic}\label{Alg-SI-Suppress}
\end{algorithm}

\begin{algorithm}[!htbp]
    \caption{Pseudo-Code for the Proposed Transceiver and RIS Configuration Design Considering RIS Self-Interference}
    \begin{algorithmic}[1]
        \State Initialize $\bm b$ and $\bm \Psi \triangleq (\bm I + \bm \Phi \bm H) \bm \Phi$ to ensure that the power constraints in \eqref{OP1-NodePowerCons} and \eqref{OP1-RISPowerCons} are satisfied (replacing $\bm \Phi$ with $\bm \Psi$);
        \While{not converge}
        \State Update $\bm m$ through \eqref{Optimal-m} (replacing $\bm \Phi$ with $\bm \Psi$);
        \State Update $\bm b$ through solving \eqref{Opt-b} (replacing $\bm \Phi$ with $\bm \Psi$);
        \State Update $\bm \Phi$ through \textbf{Algorithm \ref{Alg-SI-Suppress}}.
        \EndWhile
    \end{algorithmic}\label{Alg-SI}
\end{algorithm}

\section{Numerical Results}\label{Sec-NR}
In this section, we present numerical results to demonstrate the effectiveness of deploying an active RIS in minimizing the MSE for the considered AirComp system.

\subsection{Simulation Settings}
We adopt a three-dimensional coordinate configuration, where the locations of the AP and the active RIS are set to $\left(-50, 0, 10\right)$ meters and $\left(0, 0, 10\right)$ meters, respectively, and the $K$ users are uniformly distributed in the region of $\left([0,20], [-10, 10], 0\right)$ meters. Each link in $\{\bm h_{r,k}\}$, $\{\bm h_{d,k}\}$, and $\bm G$ is subjected to both path loss and small-scale fading. The path loss model is given by $\textsf{PL}(\delta) = T_0 \left(\delta / \delta_0\right)^{-\beta}$, where $T_0 = 30$ dB denotes the path loss at the reference distance of $\delta_0 = 1$ meter, $\delta$ represents the link distance, and $\beta$ is the path loss component. Throughout the simulations, the path loss components for $\{\bm h_{r,k}\}$, $\{\bm h_{d,k}\}$, and $\bm G$ are set to $2.8$, $3.6$, and $2.2$, respectively. For small-scale fading, we employ the standard Rician channel model, assigning Rician factors of $0$, $0$, and $3$ dB to $\{\bm h_{r,k}\}$, $\{\bm h_{d,k}\}$, and $\bm G$, respectively. Furthermore, we set $P_r = 0$ dB, $P_k = 0$ dB, $\forall k \in \mathcal K$, and $\sigma_a^2 = \sigma_r^2 = \sigma_0^2$.

\subsection{Results for Ideal RIS Model}

\begin{figure}
\centering
\includegraphics[width = 7.8cm]{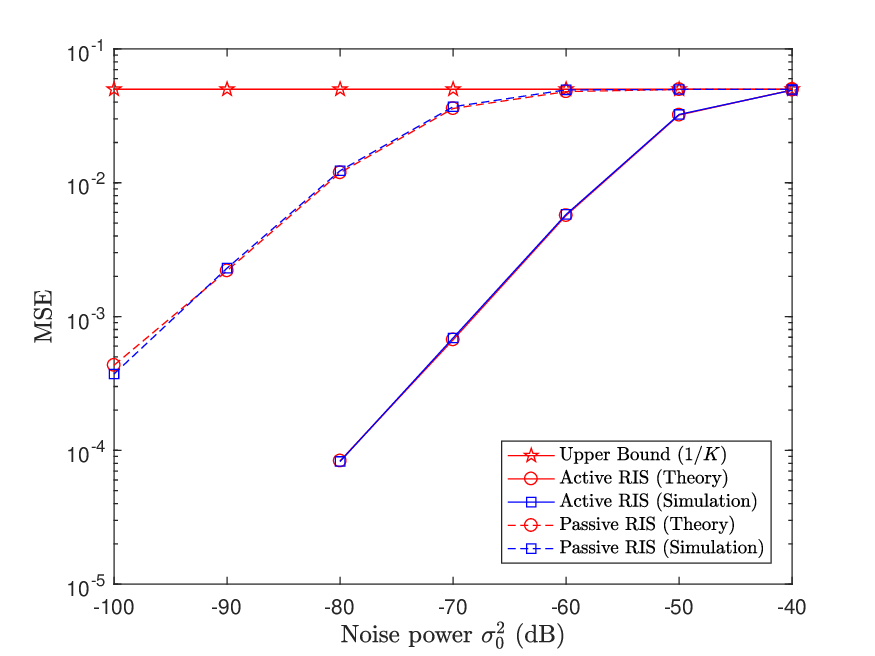}
\caption{MSE versus the noise power $\sigma_0^2$, where $M = 10$, $K = 20$, and $N = 200$.}\label{Fig-MSE-SNR}
\end{figure}

Fig. \ref{Fig-MSE-SNR} shows the effects of noise power $\sigma_0^2$ on the MSE of the active RIS-aided AirComp system. As a benchmark, the MSE of the passive RIS-aided system is also provided. An immediate observation from this figure is the notable superiority of the active RIS over its passive counterpart in reducing the MSE of the AirComp system when $\sigma_0^2$ remains low, demonstrating the advantages of incorporating active RISs into AirComp systems. Moreover, as $\sigma_0^2$ increases, the MSE of the active RIS-aided AirComp system consistently degrades until reaching its upper bound $1/K$, as given in \eqref{Eq-MSE2}. We also observe the congruence between our derived MSE in \eqref{Eq-MSE} and the simulation outcomes.

\subsection{Asymptotic Results}

\begin{figure}
\centering
\includegraphics[width = 7.8cm]{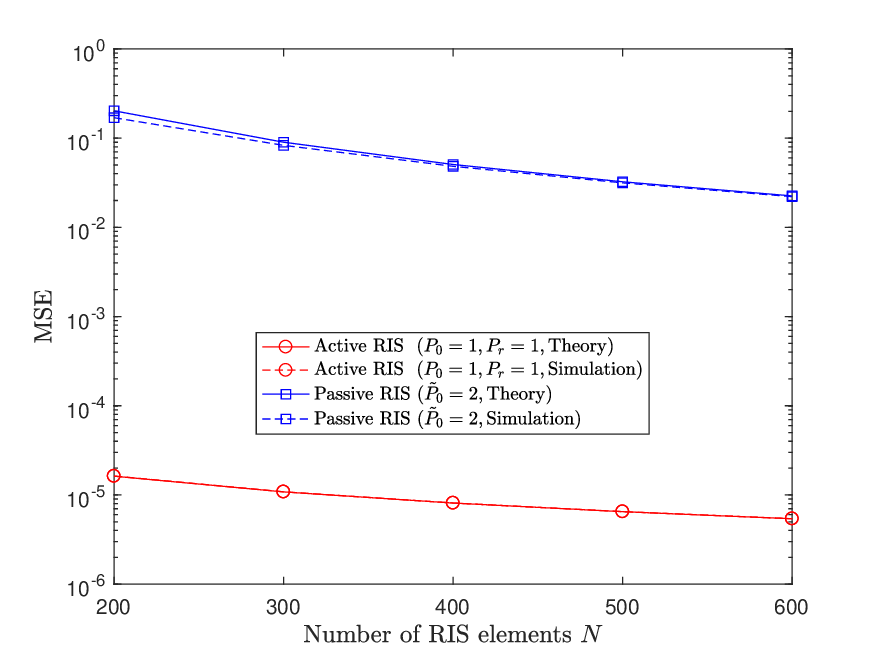}
\caption{MSE versus the number of RIS elements $N$, where $M = 1$, $K = 1$, $\rho_g^2 = \rho_r^2 = -70$ dB, and $\sigma_0^2 = -100$ dB.}\label{Fig-MSE-N1}
\end{figure}

Fig. \ref{Fig-MSE-N1} shows the influence of varying RIS elements $N$ on the MSE for the SU-SISO scenario. From this figure, we readily observe that our derived asymptotic MSEs for both the active RIS-aided and passive RIS-aided systems (referenced in \eqref{MSE-SISO-ActiveRIS} and \eqref{MSE-SISO-PassiveRIS}, respectively) coincide well with their respective simulation outcomes. Moreover, as $N$ increases, the MSEs for both systems decline. The reason is straightforward: an RIS with more elements, whether active or passive, offers more flexibility in the design of its beamforming matrix.

\begin{figure}
\centering
\includegraphics[width = 7.8cm]{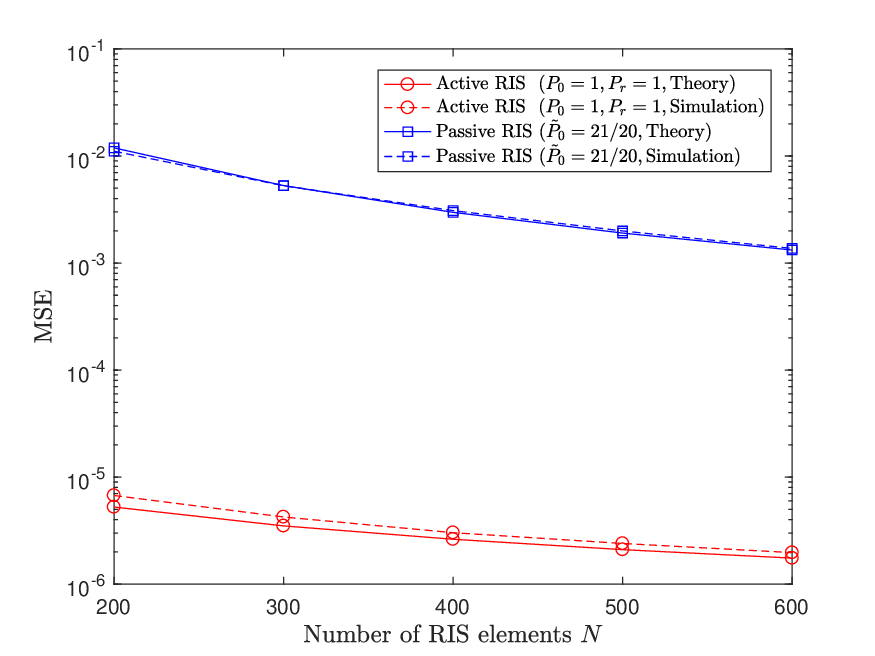}
\caption{MSE versus the number of RIS elements $N$, where $M = N$, $K = 20$, $\rho_g^2 = \rho_r^2 = -70$ dB, and $\sigma_0^2 = -100$ dB.}\label{Fig-MSE-N2}
\end{figure}

On the other hand, the impact of $N$ on the MSE for the MU-SIMO scenario is presented in Fig. \ref{Fig-MSE-N2}. Again, we observe that our derived asymptotic MSEs for both the active RIS-aided and passive RIS-aided systems (referenced in \eqref{MSE-SIMO-ActiveRIS} and \eqref{MSE-SIMO-PassiveRIS}, respectively) coincide well with their corresponding simulation outcomes. Moreover, as $N$ increases, the MSEs for both systems also decline.

\subsection{Results for Self-Interference Suppression}
In this subsection, we present numerical results to verify the effectiveness of our proposed transceiver and RIS configuration design for scenarios involving self-interference of the active RIS.

\begin{figure}
\vskip-8pt
\centering
\includegraphics[width = 7.8cm]{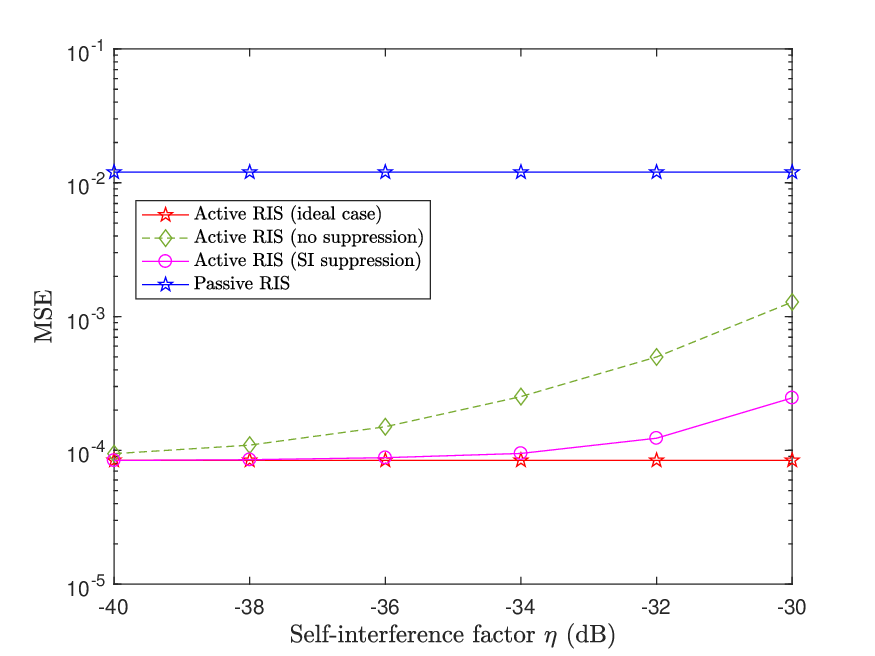}
\caption{MSE versus the self-interference factor $\eta$, where $M = 10$, $K = 20$, $N = 200$, and $\sigma_0^2 = -80$ dB.}\label{Fig-MSE-SI}
\end{figure}

As in \cite{LinglongDai-ActiveRIS}, we assume each element of the self-interference channel $\bm H$ follows a Rayleigh distribution, i.e., $\mathcal{CN}(0, \eta^2)$, where $\eta$ is termed the self-interference factor. Fig. \ref{Fig-MSE-SI} illustrates the impact of $\eta$ on the MSE for the active RIS-aided AirComp system. From this figure, we observe that without self-interference suppression, the performance of the system degrades heavily as $\eta$ increases from $-40$ dB to $-30$ dB. In contrast, by employing our proposed transceiver and RIS configuration design to counteract the self-interference, the active RIS-aided AirComp system maintains a commendable performance.

\section{Conclusions}\label{Sec-CN}
In this paper, we have introduced a novel RIS architecture, termed the active RIS, to mitigate AirComp errors, paving the way for fast and accurate WDA. Firstly, we considered the ideal RIS model and formulated an MSE minimization problem, aiming to jointly optimize the transceiver design and RIS configuration. We then developed an AO framework to solve this problem with a convergence guarantee. Subsequently, we investigated two specific system settings and analyzed their respective asymptotic MSE to reveal the advantage of incorporating active RISs into AirComp systems. Furthermore, we extended our transceiver and RIS configuration design to accommodate the self-interference of the active RIS. Simulation results showed that the active RIS significantly outperformed its passive counterpart in reducing AirComp errors.

\begin{appendices}
\section{}\label{App-Opt-Phi}
First of all, we rewrite \eqref{Opt-Phi-Obj} as a function of $\bm \phi$, given by
\begin{align*}
    f_0(\bm \phi) \overset{(a)}{=} & \sum\limits_{k=1}^K \left|\bm m^H \bm G \bm \Phi \bm h_{r, k} b_k + \bm m^{H} \bm h_{d, k} b_k - \frac{1}{K}\right|^2 \\[3ex] 
    & + \sigma_r^2 \|\bm \Phi \bm G^H \bm m\|^2 \\[3ex]
    \overset{(b)}{=} & \sum\limits_{k=1}^K \left|\bm m^H \bm G {\rm diag}(\bm h_{r, k}) \bm \phi b_k + \bm m^{H} \bm h_{d, k} b_k - \frac{1}{K}\right|^2 \\[3ex] 
    & + \sigma_r^2 \|{\rm diag}(\bm G^H \bm m) \bm \phi\|^2 \\[2ex]
    = &~ \bm \phi^H \left(\sum\limits_{k=1}^K \bm a_k \bm a_k^H  + \sigma_r^2 {\rm diag}(\bm m^H \bm G) {\rm diag}(\bm G^H \bm m) \right)\bm \phi \\[2ex]
    & + \sum\limits_{k=1}^K |c_k|^2 - \bm \phi^H \left(\sum\limits_{k=1}^K c_k \bm a_k\right) - \left(\sum\limits_{k=1}^K c^H_k \bm a_k^H\right) \bm \phi \\[2ex]
    = &~ \bm \phi^H \bm A_0 \bm \phi - \bm \phi^H \bm v_0 - \bm v_0^H \bm \phi + \sum\limits_{k=1}^K |c_k|^2,
\end{align*}
where $(a)$ is due to the property of $\bm \Phi^H \bm \Phi = \bm \Phi \bm \Phi^H$, $(b)$ is due to $\bm \Phi \bm x = {\rm diag}(\bm x) \bm \phi$, and $c_k = \frac{1}{K} - \bm m^{H} \bm h_{d, k} b_k$. Then, we write \eqref{Opt-Phi-PowerCons} as follows:
\begin{align*}
    g_0(\bm \phi) \overset{(a)}{=} & \sum\limits_{k=1}^K |b_k|^2 \|{\rm diag}(\bm h_{r, k}) \bm \phi\|^2 + \sigma_r^2 \|\bm \phi\|^2 \\[2ex]
    = &~ \bm \phi^H \left(\sum\limits_{k = 1}^K |b_k|^2 {\rm diag}(\bm h^*_{r, k}) {\rm diag}(\bm h_{r, k}) + \sigma_r^2 \bm I\right) \bm \phi \\[2ex]
    = &~ \bm \phi^H \bm B_0 \bm \phi,
\end{align*}
where $(a)$ is due to ${\rm Tr}(\bm \Phi \bm \Phi^H) = \|\bm \Phi\|_{\rm F}^2 = \|\bm \phi\|^2$.

\section{}\label{Analysis-SU-SISO}
From \eqref{Eq-MSE2}, we can derive the MSE corresponding to the case of $K = M = 1$ as follows:
\begin{align}\label{SISO-MSE1}
    & \mathbb{MSE} = \frac{\sigma_a^2 + \sigma_r^2 \|\bm \Phi \bm g\|^2}{\sigma_a^2 + \sigma_r^2 \|\bm \Phi \bm g\|^2 + |b|^2 |\bm g^H \bm \Phi \bm h_r|^2} \nonumber \\[3ex]
    &~~ \overset{(a)}{=} \frac{\sigma_a^2 + \alpha^2 \sigma_r^2 \|\bm g\|^2}{\sigma_a^2 + \alpha^2 \sigma_r^2 \|\bm g\|^2 + \alpha^2 |b|^2 \left|\sum\nolimits_{n=1}^N e^{j \omega_n} g_n^* h_{r,n}\right|^2},
\end{align}
where $(a)$ is due to $\|\bm \Phi \bm g\|^2 = \alpha^2 \|\bm g\|^2$, and $g_n$, $h_{r,n}$ represent the $n$-th entry in $\bm g$ and $\bm h_r$, respectively. To minimize \eqref{SISO-MSE1}, we set $b$, $\omega_n$, $\forall n \in \mathcal N$, and $\alpha$ as follows:
\begin{subequations}\label{SISO-Optimal-Sols}
\begin{align}
    & b^{\star} = \sqrt{P_0}, \\[3ex]
    & \omega_n^{\star} = \angle g_n - \angle h_{r,n}, \\[2ex]
    & \alpha^{\star} = \sqrt{\frac{P_r}{P_0 \|\bm h_r\|^2 + N \sigma_r^2}},
\end{align}    
\end{subequations}
where $\angle g_n$ and $\angle h_{r,n}$ denote the phases of $g_n$ and $h_{r,n}$, respectively. By substituting \eqref{SISO-Optimal-Sols} into \eqref{SISO-MSE1}, we obtain that
\begin{align}\label{SISO-MSE2}
    \mathbb{MSE} & = \frac{(P_0 \|\bm h_r\|^2 + N \sigma_r^2) \sigma_a^2 + P_r \sigma_r^2 \|\bm g\|^2}{(P_0 \|\bm h_r\|^2 + N \sigma_r^2) \sigma_a^2 + P_r \sigma_r^2 \|\bm g\|^2 + P_r P_0 e^2} \nonumber \\[2ex]
    & \overset{(a)}{\approx} \frac{1}{N} \frac{(P_0 \rho^2_r + \sigma_r^2) \sigma_a^2 + P_r \sigma_r^2 \rho_g^2}{\frac{(P_0 \rho^2_r + \sigma_r^2) \sigma_a^2  + P_r \sigma_r^2 \rho^2_g}{N} + P_r P_0 \frac{\pi^2 \rho^2_r \rho_g^2}{16}} \nonumber \\[3ex]
    & \approx \frac{16}{N} \frac{(P_0 \rho^2_r + \sigma_r^2) \sigma_a^2 + P_r \sigma_r^2 \rho_g^2}{P_r P_0 \pi^2 \rho^2_r \rho_g^2},
\end{align}
where $e = \sum\nolimits_{n=1}^N |g_n| |h_{r,n}|$, and $(a)$ is due to the following properties: $\lim\nolimits_{N \to \infty} \frac{1}{N} \|\bm h_r\|^2 = \rho^2_r$, $\lim\nolimits_{N \to \infty} \frac{1}{N} \|\bm g\|^2 = \rho^2_g$, and $\lim\nolimits_{N \to \infty} \frac{1}{N} \sum\nolimits_{n=1}^N |g_n| |h_{r,n}| = \frac{\pi \rho_r \rho_g}{4}$. Until now, we have proved Theorem \ref{Theorem-SISO-ActiveRIS}.

\section{}\label{Analysis-SU-SISO2}
By letting $\sigma_r^2 = 0$ in \eqref{SISO-MSE1}, we obtain the MSE for the passive RIS-aided SU-SISO system:
\begin{align}\label{SISO-MSE3}
    \mathbb{MSE} & = \frac{\sigma_a^2}{\sigma_a^2 + |\tilde b|^2 |\bm g^H \breve{\bm \Phi} \bm h_r|^2} \nonumber \\[2ex]
    & = \frac{\sigma_a^2}{\sigma_a^2 + |\tilde b|^2 \left|\sum\nolimits_{n=1}^N e^{j \omega_n} g_n^{*} h_{r,n} \right|^2},
\end{align}
where $\breve{\bm \Phi} = {\rm diag}(e^{j \omega_1}, \cdots, e^{j \omega_N})$ is the beamforming matrix of the passive RIS. To minimize \eqref{SISO-MSE3}, we set $\tilde b$ and $\omega_n$, $\forall n \in \mathcal N$, as follows:
\begin{subequations}\label{SISO-Optimal-Sols2}
\begin{align}
    & \tilde b^{\star} = \sqrt{\tilde P_0}, \\[2ex]
    & \omega_n^{\star} = \angle g_n - \angle h_{r,n}.
\end{align}    
\end{subequations}
Lastly, by substituting \eqref{SISO-Optimal-Sols2} into \eqref{SISO-MSE3}, we have
\begin{align}\label{SISO-MSE4}
    \mathbb{MSE} & = \frac{\sigma_a^2}{\sigma_a^2 + \tilde P_0 \left(\sum\nolimits_{n=1}^N |g_n| |h_{r,n}|\right)^2} \nonumber \\[2ex]
    & = \frac{1}{N^2} \frac{\sigma_a^2}{\sigma_a^2 / N^2 + \tilde P_0 \left(\sum\nolimits_{n=1}^N |g_n| |h_{r,n}| / N\right)^2} \nonumber \\[2ex]
    & \approx \frac{16}{N^2} \frac{\sigma_a^2}{\tilde P_0 \pi^2 \rho^2_r \rho_g^2},
\end{align}
which proves Theorem \ref{Theorem-SISO-PassiveRIS}.

\section{}\label{Analysis-MU-SIMO}
It is observed from \eqref{Eq-MSE2} that the main challenge in analyzing the computed MSE is $\bm R^{-1}$, which, however, can be notably simplified using \eqref{Channel-Property2}, as detailed below.

First of all, we employ the property of $\bm \Phi \bm \Phi^H = \alpha^2 \bm I$ and $\bm G \bm G^H / N \to \rho_g^2 \bm I$ to simplify $\bar{\bm R} = \bm R / N^2$ as follows:
\begin{align}\label{Analysis-SIMO-Eq1}
    \bar{\bm R} & = \frac{1}{N^2} \left(\sum\limits_{k=1}^K |b_k|^2 \bm h_{e, k} \bm h_{e,k}^H + \sigma_r^2 \bm G \bm \Phi \bm \Phi^{H} \bm G^H + \sigma_a^2 \bm I\right) \nonumber \\[2ex]
    & \approx \sum\limits_{k=1}^K |b_k|^2 \bar{\bm h}_{e, k} \bar{\bm h}_{e,k}^H + \left(\frac{\alpha^2 \rho_g^2 \sigma_r^2}{N}  + \frac{\sigma_a^2}{N^2}\right) \bm I,
\end{align}
where $\bar{\bm h}_{e,k} = \bm h_{e,k} / N$, $\forall k \in \mathcal K$. Next, we define the following matrix sequence:
\begin{equation}\label{Analysis-SIMO-Eq2}
    \bm D_k = \sum\limits_{j=1}^k |b_j|^2 \bar{\bm h}_{e,j} \bar{\bm h}_{e,j}^H + \sigma^2 \bm I, ~\forall k = 1, \cdots, K, 
\end{equation}
where $\sigma^2 = {\alpha^2 \rho_g^2 \sigma_r^2}/{N} + {\sigma_a^2}/{N^2}$. It is not difficult to see that
\begin{equation}\label{Analysis-SIMO-Eq3}
    \bm D_k = \bm D_{k-1} + |b_k|^2 \bar{\bm h}_{e,k} \bar{\bm h}_{e,k}^H, ~\forall k = 2, \cdots, K,
\end{equation}
and $\bm D_K = \bar{\bm R}$. According to \eqref{Analysis-SIMO-Eq3} and the Sherman-Morrison formula, we can compute $\bm D^{-1}_k$, $\forall k = 2, \cdots, K$, in a recursive manner:
\begin{equation}\label{Analysis-SIMO-Eq4}
    \bm D^{-1}_k = \bm D^{-1}_{k-1} - \frac{|b_k|^2 \|\bm D^{-1}_{k-1} \bar{\bm h}_{e,k}\|^2}{1 + |b_k|^2 \bar{\bm h}^H_{e,k} \bm D^{-1}_{k-1} \bar{\bm h}_{e,k}}.
\end{equation}
Note that $\bm D^{-1}_1$ can be computed as follows:
\begin{equation}\label{Analysis-SIMO-Eq5}
    \bm D^{-1}_1 = \frac{1}{\sigma^2} \left(\bm I - \frac{|b_1|^2 \bar{\bm h}_{e,1} \bar{\bm h}_{e,1}^H}{\sigma^2 + |b_1|^2 \|\bar{\bm h}_{e,1}\|^2}\right).
\end{equation}
Let $k = 2$, and by substituting \eqref{Analysis-SIMO-Eq5} into \eqref{Analysis-SIMO-Eq4}, we have
\begin{align*}
    & \bm D^{-1}_2 = \bm D^{-1}_1 - \frac{|b_2|^2 \|\bm D^{-1}_1 \bar{\bm h}_{e,2}\|^2}{1 + |b_2|^2 \bar{\bm h}^H_{e,2} \bm D^{-1}_1 \bar{\bm h}_{e,2}}, \\[2ex]
    &~~ \overset{(a)}{\approx} \frac{1}{\sigma^2} \left(\bm I - \frac{|b_1|^2 \bar{\bm h}_{e,1} \bar{\bm h}_{e,1}^H}{\sigma^2 + |b_1|^2 \|\bar{\bm h}_{e,1}\|^2} - \frac{|b_2|^2 \bar{\bm h}_{e,2} \bar{\bm h}_{e,2}^H}{\sigma^2 + |b_2|^2 \|\bar{\bm h}_{e,2}\|^2}\right),
\end{align*}
where $(a)$ follows due to $\bar{\bm h}^H_{e,1} \bar{\bm h}_{e,2} = 0$. By recursively computing $\bm D^{-1}_3, \cdots, \bm D^{-1}_K$, we obtain that
\begin{equation}\label{Analysis-SIMO-Eq6}
    \bar{\bm R}^{-1} = \bm D^{-1}_K \approx \frac{1}{\sigma^2} \left(\bm I - \sum\limits_{k=1}^K \frac{|b_k|^2 \bar{\bm h}_{e,k} \bar{\bm h}_{e,k}^H}{\sigma^2 + |b_k|^2 \|\bar{\bm h}_{e,k}\|^2}\right).
\end{equation}
Lastly, by substituting \eqref{Analysis-SIMO-Eq6} into \eqref{Eq-MSE2}, we have
\begin{align}\label{Analysis-SIMO-Eq7}
    \mathbb{MSE} & = \frac{1}{K} - \frac{1}{K^2} \left(\sum\limits_{k=1}^K \bar{\bm h}^H_{e, k} b^*_k\right) \bar{\bm R}^{-1} \left(\sum\limits_{k=1}^K \bar{\bm h}_{e, k} b_k\right) \nonumber \\[2ex]
    & \approx \frac{1}{K} - \frac{1}{K^2 \sigma^2} \left(\sum\limits_{k=1}^K \bar{\bm h}^H_{e, k} b^*_k\right) \nonumber \\[2ex]
    & \times \left(\bm I - \sum\limits_{j=1}^K \frac{|b_j|^2 \bar{\bm h}_{e,j} \bar{\bm h}_{e,j}^H}{\sigma^2 + |b_j|^2 \|\bar{\bm h}_{e,j}\|^2}\right) \left(\sum\limits_{l=1}^K \bar{\bm h}_{e, l} b_l\right) \nonumber \\[2ex]
    & \overset{(a)}{\approx} \frac{1}{K} - \frac{1}{K^2} \sum\limits_{k=1}^K \frac{|b_k|^2 \|\bar{\bm h}_{e,k}\|^2}{\sigma^2 + |b_k|^2 \|\bar{\bm h}_{e,k}\|^2} \nonumber \\[2ex]
    & \overset{(b)}{\approx} \frac{1}{K} - \frac{1}{K^2} \sum\limits_{k=1}^K \frac{|b_k|^2 \alpha^2 \rho^2_g \rho^2_r}{\sigma^2 + |b_k|^2 \alpha^2 \rho^2_g \rho^2_r} \nonumber \\[2ex]
    & \overset{(c)}{\approx} \frac{1}{K N} \frac{\sigma_r^2 \sigma_a^2 + K P_0 \rho^2_r \sigma_a^2 + P_r \rho^2_g \sigma_r^2}{P_r P_0 \rho_r^2 \rho_g^2},
\end{align}
where $(a)$ follows due to $\bar{\bm h}^H_{e,k} \bar{\bm h}_{e,k'} = 0$, $\forall k, k' \in \mathcal K$, $(b)$ follows due to $\|\bar{\bm h}_{e,k}\|^2 = \alpha^2 \rho^2_g \rho^2_r$, $\forall k \in \mathcal K$, and $(c)$ follows since we set $|b_k|^2 = P_0$, $\forall k \in {\cal K}$, and $\alpha^2 = \frac{P_r}{N(\sigma_r^2 + K P_0 \rho^2_r)}$ to meet the power constraints of each user and the active RIS. Until now, Theorem \ref{Theorem-SIMO-ActiveRIS} is proved.

\section{}\label{Analysis-MU-SIMO2}
By letting $\sigma_r^2 = 0$ in $\bm R$, we obtain the MSE for the passive RIS-aided MU-SIMO system:
\begin{equation}\label{Analysis-SIMO-Eq8}
    \mathbb{MSE} = \frac{1}{K} - \frac{1}{K^2} \left(\sum\limits_{k=1}^K \tilde{\bm h}^H_{e, k} \tilde b^*_k\right) \tilde{\bm R}^{-1} \left(\sum\limits_{k=1}^K \tilde{\bm h}_{e, k} \tilde b_k\right),
\end{equation}
where $\tilde{\bm h}_{e, k} = \bm G \breve{\bm \Phi} \bm h_{r,k} / N$, $\forall k \in \mathcal K$, and $\tilde{\bm R}$ is given by
\begin{equation}
    \tilde{\bm R} = \sum\limits_{k=1}^K |\tilde b_k|^2 \tilde{\bm h}_{e, k} \tilde{\bm h}_{e,k}^H + \frac{\sigma_a^2}{N^2} \bm I.
\end{equation}
Similar to $\bar{\bm R}^{-1}$ in \eqref{Analysis-SIMO-Eq6}, we approximate $\tilde{\bm R}^{-1}$ as follows:
\begin{equation}\label{Analysis-SIMO-Eq9}
    \tilde{\bm R}^{-1} \approx \frac{1}{\tilde \sigma^2} \left(\bm I - \sum\limits_{k=1}^K \frac{|\tilde b_k|^2 \tilde{\bm h}_{e,k} \tilde{\bm h}_{e,k}^H}{\tilde \sigma^2 + |\tilde b_k|^2 \|\tilde{\bm h}_{e,k}\|^2}\right),
\end{equation}
where $\tilde \sigma^2 = \sigma_a^2 / N^2$. By substituting \eqref{Analysis-SIMO-Eq9} into \eqref{Analysis-SIMO-Eq8}, we have
\begin{align}
    \mathbb{MSE} & \approx \frac{1}{K} - \frac{1}{K^2 \tilde \sigma^2} \left(\sum\limits_{k=1}^K \tilde{\bm h}^H_{e, k} \tilde b^*_k\right) \nonumber \\[2ex]
    & \times \left(\bm I - \sum\limits_{j=1}^K \frac{|\tilde b_j|^2 \tilde{\bm h}_{e,j} \tilde{\bm h}_{e,j}^H}{\tilde \sigma^2 + |\tilde b_j|^2 \|\tilde{\bm h}_{e,j}\|^2}\right) \left(\sum\limits_{l=1}^K \tilde{\bm h}_{e, l} \tilde b_l\right) \nonumber \\[2ex]
    & \overset{(a)}{\approx} \frac{1}{K} - \frac{1}{K^2} \sum\limits_{k=1}^K \frac{|\tilde b_k|^2 \|\tilde{\bm h}_{e,k}\|^2}{\tilde \sigma^2 + |\tilde b_k|^2 \|\tilde{\bm h}_{e,k}\|^2} \nonumber \\[2ex]
    & \overset{(b)}{\approx} \frac{1}{K} - \frac{1}{K^2} \sum\limits_{k=1}^K \frac{|\tilde b_k|^2 \rho^2_g \rho^2_r}{\tilde \sigma^2 + |\tilde b_k|^2 \rho^2_g \rho^2_r} \nonumber \\[2ex]
    & \overset{(c)}{\approx} \frac{1}{K N^2} \frac{\sigma_a^2}{\tilde P_0 \rho_r^2 \rho_g^2},
\end{align}
where $(a)$ follows due to $\tilde{\bm h}^H_{e,k} \tilde{\bm h}_{e,k'} = 0$, $\forall k, k' \in \mathcal K$, $(b)$ follows due to $\|\tilde{\bm h}_{e,k}\|^2 = \rho^2_g \rho^2_r$, $\forall k \in \mathcal K$, and $(c)$ follows by setting $|\tilde b_k|^2 = \tilde P_0$, $\forall k \in \mathcal K$. Thus far, we have proved Theorem \ref{Theorem-SIMO-PassiveRIS}.

\section{}\label{App-Opt-Theta}
First of all, we rewrite \eqref{Opt-Theta2-Obj} as a function of $\bm \phi$, given by
\begin{align*}
    & f_3(\bm \phi) = \sum\limits_{k=1}^K \left|\bm m^{H} (\bm h_{d, k} + \bm G \bm \Omega \bm \Phi \bm h_{r, k}) b_k - \frac{1}{K}\right|^2 \\[2ex]
    &~~~~~~~~~~ + \sigma_r^2 \|\bm \Phi \bm \Omega^H \bm G^H \bm m\|^2 + \tau \|\bm {\tilde \Phi} - \bm \Phi\|_{\rm F}^2 \\[2ex]
    & = \sum\limits_{k=1}^K \left|\bm{\bar m}^H \bm H_{r, k} b_k \bm \phi - c_k \right|^2 + \sigma_r^2 \|\bm{\bar M} \bm \phi\|^2 + \tau \|\bm {\tilde \phi} - \bm \phi\|^2 \\[2ex]
    & = \bm \phi^H \left(\sum\limits_{k=1}^K \bm {\bar a}_k \bm {\bar a}_k^H  + \sigma_r^2 \bm{\bar M}^H \bm{\bar M} + \tau \bm I \right) \bm \phi + C_1 \\[2ex]
    &~~~ - \bm \phi^H \left(\tau \bm {\tilde \phi} + \sum\limits_{k=1}^K c_k \bm {\bar a}_k\right) - \left(\tau \bm {\tilde \phi}^H + \sum\limits_{k=1}^K c^H_k \bm {\bar a}_k^H\right) \bm \phi \\[2ex]
    & = \bm \phi^H \bm A_1 \bm \phi - \bm \phi^H \bm v_1 - \bm v_1^H \bm \phi + C_1,
\end{align*}
where $\bm{\bar m} = \bm \Omega^H \bm G^H \bm m$, $\bm{\bar M} = {\rm diag}(\bm{\bar m})$, $\bm H_{r,k} = {\rm diag}(\bm h_{r, k})$, and $C_1 = \sum\nolimits_{k=1}^K |c_k|^2 + \tau \|\bm {\tilde \phi}\|^2$. Subsequently, we rewrite \eqref{Opt-Theta2-PowerCons} as follows:
\begin{align*}
    g_3(\bm \Phi) = & \sum\limits_{k=1}^K |b_k|^2 \|\bm \Omega \bm \Phi \bm h_{r, k}\|^2 + \sigma_r^2 {\rm Tr}(\bm \Omega \bm \Phi \bm \Phi^H \bm \Omega^H) \\[2ex]
    \overset{(a)}{=} & \sum\limits_{k=1}^K |b_k|^2 \|\bm \Omega \bm H_{r, k} \bm \phi\|^2 + \sigma_r^2 {\rm Tr}(\bm \Omega^H \bm \Omega \bm \Phi \bm \Phi^H) \\[2ex]
    \overset{(b)}{=} &~ \bm \phi^H \Bigg(\sum\limits_{k=1}^K |b_k|^2 \bm H^*_{r,k} \bm \Omega^H \bm \Omega \bm H_{r, k} + \sigma_r^2 (\bm \Omega^H \bm \Omega \odot \bm I) \Bigg) \bm \phi \\[2ex]
    = &~ \bm \phi^H \bm B_1 \bm \phi,
\end{align*}
where $(a)$ is due to the property of ${\rm Tr}(\bm X \bm Y) = {\rm Tr}(\bm Y \bm X)$, and $(b)$ is due to the property of ${\rm Tr}(\bm X \bm \Phi \bm \Phi^H) = \bm \phi^H (\bm X \odot \bm I) \bm \phi$.

\section{}\label{App-Opt-Theta2}
First of all, we rewrite \eqref{Opt-Theta2-Obj} as a function of $\bm {\tilde \phi}$, given by
\begin{align*}
    & f_4(\bm {\tilde \phi}) = \sum\limits_{k=1}^K \left|\bm m^H (\bm h_{d, k} + \bm G (\bm I + \bm {\tilde \Phi} \bm H) \bm \Phi \bm h_{r, k}) b_k - \frac{1}{K}\right|^2 \\[2ex]
    &~~~~~~~~~~ + \sigma_r^2 \|\bm m^{H} \bm G (\bm I + \bm {\tilde \Phi} \bm H) \bm \Phi\|^2 + \tau \|\bm {\tilde \Phi} - \bm \Phi\|_{\rm F}^2 \\[2ex]
    & = \sum\limits_{k=1}^K \left|\bm {\tilde a}_k^H \bm {\tilde \phi} - {\tilde c}_k \right|^2 + \sigma_r^2 \Big( \bm m^H \bm G \bm {\tilde \Phi} \bm H \bm \Phi \bm \Phi^H \bm G^H \bm m \\[2ex]
    &~~~~ + \|\bm m^H \bm G \bm \Phi\|^2 + \bm m^H \bm G \bm \Phi \bm \Phi^H \bm H^H \bm {\tilde \Phi}^H \bm G^H \bm m \\[2ex]
    &~~~~ + \bm m^H \bm G \bm {\tilde \Phi} \bm H \bm \Phi \bm \Phi^H \bm H^H \bm {\tilde \Phi}^H \bm G^H \bm m\Big) + \tau \|\bm {\tilde \phi} - \bm \phi\|_2^2 \\[1ex]
    & \overset{(a)}{=} \sum\limits_{k=1}^K \left|\bm {\tilde a}_k^H \bm {\tilde \phi} - \tilde{c}_k \right|^2 + \tau \|\bm {\tilde \phi} - \bm \phi\|^2 \\[2ex]
    &~~~~ + \sigma_r^2 \Big(\|\bm m^H \bm G \bm \Phi\|^2 + \bm {\tilde \phi}^H \bm \Sigma_2^H \bm H^* \bm \Phi^H \bm \Phi \bm H^T \bm \Sigma_2 \bm {\tilde \phi} \\[2ex]
    &~~~~ + \bm {\tilde \phi}^H \bm \Sigma_1^H \bm G^H \bm m + \bm m^H \bm G \bm \Sigma_1 \bm {\tilde \phi} \Big) \\[2ex]
    & =~ \bm {\tilde \phi}^H \left(\sum\limits_{k=1}^K \bm {\tilde a}_k \bm {\tilde a}_k^H  + \sigma_r^2 \bm \Sigma_2^H \bm H^{*} \bm \Phi^H \bm \Phi \bm H^T \bm \Sigma_2 + \tau \bm I\right) \bm {\tilde \phi} \\[2ex]
    &~~~~ - \bm {\tilde \phi}^H \left(\sum\limits_{k=1}^K \tilde c_k \bm {\tilde a}_k + \tau \bm \phi - \sigma_r^2 \bm \Sigma_1^H \bm G^H \bm m \right) \\[2ex] 
    &~~~~ - \left(\sum\limits_{k=1}^K \tilde c^H_k \bm {\tilde a}_k^H + \tau \bm \phi^H - \sigma_r^2 \bm m^H \bm G \bm \Sigma_1 \right) \bm {\tilde \phi} + C_2 \\[2ex]
    & =~ \bm {\tilde \phi}^H \bm A_2 \bm {\tilde \phi} - \bm {\tilde \phi}^H \bm v_2 - \bm v_2^H \bm {\tilde \phi} + C_2,
\end{align*}
where $\bm \Sigma_1 = {\rm diag} (\bm H \bm \Phi \bm \Phi^H \bm G^H \bm m)$, $\bm \Sigma_2 = {\rm diag}(\bm m^H \bm G)$, $\tilde c_k = K^{-1} - \bm m^{H} (\bm h_{d, k} + \bm G \bm \Phi \bm h_{r,k}) b_k$, $C_2 = \sum\nolimits_{k=1}^K |\tilde c_k|^2 + \tau \|\bm \phi\|^2 + \sigma_r^2 \|\bm m^H \bm G \bm \Phi\|^2$, and $(a)$ is due to the following manipulations:
\begin{align*}
    & \bm m^H \bm G \bm {\tilde \Phi} \bm H \bm \Phi \bm \Phi^H \bm H^H \bm {\tilde \Phi}^H \bm G^H \bm m \\[2ex]
    &~~~ = \bm {\tilde \phi}^T {\rm diag}(\bm m^H \bm G) \bm H \bm \Phi \bm \Phi^H \bm H^H {\rm diag}(\bm G^H \bm m) \bm {\tilde \phi}^* \\[2ex]
    &~~~ = \bm {\tilde \phi}^H {\rm diag}(\bm G^H \bm m) \bm H^* \bm \Phi^* \bm \Phi \bm H^T {\rm diag}(\bm m^H \bm G) \bm {\tilde \phi}.
\end{align*}
Subsequently, we rewrite \eqref{Opt-Theta2-PowerCons} as follows:
\begin{align*}
    g_4(\bm {\tilde \phi}) = & \sum\limits_{k=1}^K |b_k|^2 {\rm Tr}((\bm I + \bm {\tilde \Phi} \bm H) \bm \Phi \bm h_{r, k} \bm h^H_{r,k} \bm \Phi^H (\bm I + \bm {\tilde \Phi} \bm H)^H) \\[1ex]
    & + \sigma_r^2 {\rm Tr}((\bm I + \bm {\tilde \Phi} \bm H) \bm \Phi \bm \Phi^H (\bm I + \bm {\tilde \Phi} \bm H)^H) \\[2ex]
    = &~ {\rm Tr}((\bm I + \bm {\tilde \Phi} \bm H) \bm D (\bm I + \bm {\tilde \Phi} \bm H)^H) \\[1ex]
    \overset{(b)}{=} &~ {\rm Tr}(\bm D + \bm {\tilde \Phi} \bm H \bm D + \bm D \bm H^H \bm {\tilde \Phi}^H + \bm {\tilde \Phi} \bm H \bm D \bm H^H \bm {\tilde \Phi}^H) \\[2ex]
    = &~ \bm {\tilde \phi}^H \bm B_2 \bm {\tilde \phi} + \bm {\tilde \phi}^H \bm q + \bm q^H \bm {\tilde \phi} + {\rm Tr}(\bm D),
\end{align*}
where $(b)$ is due to the property of ${\rm diag}(\bm X \bm {\tilde \Phi}^H) = \bm {\tilde \phi}^H \bm x$, and the property of ${\rm Tr}(\bm X \bm {\tilde \Phi}^H \bm {\tilde \Phi}) = \bm {\tilde \phi}^H (\bm X \odot \bm I) \bm {\tilde \phi}$.
\end{appendices}

\end{document}